\documentclass[runningheads]{llncs}
\usepackage[T1]{fontenc}
\usepackage{amsmath}
\usepackage{graphicx}
\usepackage{booktabs}
\usepackage{refcount}  
\usepackage{cite}
\usepackage{hyperref}
\usepackage{color}
\usepackage{multirow}
\usepackage{enumitem}
\usepackage[labelfont=bf]{subcaption}
\usepackage{float}
\usepackage{wrapfig}
\usepackage{pifont}
\usepackage{tikz}
\usepackage[ruled,lined,linesnumbered]{algorithm2e}
\usetikzlibrary{arrows.meta,positioning,calc,decorations.pathreplacing,fit}
\setlist[itemize]{label=$\bullet$}

\begin{document}
\title{Can Drift-Adaptive Malware Detectors Be Made Robust? Attacks and Defenses Under White-Box and Black-Box Threats}
\titlerunning{~}
%
\author{Adrian Shuai Li \and
Md Ajwad Akil \and
Elisa Bertino}
\authorrunning{~}
%
\institute{Purdue University\\
\email{\{li3944,makil,bertino\}@purdue.edu}}


\maketitle              
\begin{abstract}
Concept drift and adversarial evasion are two major challenges for deploying machine learning-based malware detectors. While both have been studied separately, their combination, the adversarial robustness of drift-adaptive detectors, remains unexplored. We address this problem with AdvDA, a recent malware detector that uses adversarial domain adaptation to align a labeled source domain with a target domain with limited labels. The distribution shift between domains poses a unique challenge: robustness learned on the source may not transfer to the target, and existing defenses assume a fixed distribution. To address this, we propose a universal robustification framework that fine-tunes a pretrained AdvDA model on adversarially transformed inputs, agnostic to the attack type and choice of transformations. We instantiate it with five defense variants spanning two threat models: white-box PGD attacks in the feature space and black-box MalGuise attacks that modify malware binaries via functionality-preserving control-flow mutations. Across nine defense configurations, five monthly adaptation windows on Windows malware, and three false-positive-rate operating points, we find the undefended AdvDA completely vulnerable to PGD (100\% attack success) and moderately to MalGuise (13\%). Our framework reduces these rates to as low as 3.2\% and 5.1\%, respectively, but the optimal strategy differs: source adversarial training is essential for PGD defenses yet counterproductive for MalGuise defenses, where target-only training suffices. Furthermore, robustness does not transfer across these two threat models. We provide deployment recommendations that balance robustness, detection accuracy, and computational cost.

\end{abstract}
\section{Introduction}

Malware remains one of the most persistent threats to computer security, with hundreds of thousands of new samples appearing daily. Machine learning-based detectors are now widely deployed to cope with this volume, but they face a fundamental challenge: malware authors continuously evolve their techniques, causing a detector trained on past samples to gradually lose its effectiveness on newer threats. This phenomenon is known as concept drift~\cite{chen2023continuous,yang2021cade,barbero2022transcending,jordaney2017transcend,ma2021comprehensive,ceschin2023fast,ceschin2024machine}.

A recent drift-adaptive detector, AdvDA~\cite{li2025revisiting}, addresses this through adversarial domain adaptation (DA): it periodically realigns feature distributions between older, labeled malware samples (the \emph{source} domain) and newer samples with few labels (the \emph{target} domain), maintaining strong detection accuracy as the threat landscape shifts. However, AdvDA, like other drift-adaptation methods, has been designed and evaluated only for clean detection accuracy. In practice, malware authors are active adversaries who craft samples to evade detection~\cite{kreuk2018adversarial,lucas2021malware,ling2024wolf,zhang2022semantics}, and if an attacker can reliably evade a drift-adapted model, the entire adaptation pipeline is undermined regardless of its clean performance.

Yet protecting AdvDA against adversarial attacks poses unique challenges. Existing defenses~\cite{chakraborty2018adversarial,ren2020adversarial} for adversarial perturbations~\cite{szegedy2013intriguing,goodfellow2014explaining,chakraborty2018adversarial,hendrycks2019benchmarking} assume a fixed data distribution and do not address the distribution shift inherent in AdvDA, where robustness learned on the source domain may not transfer to the target. To address this, we propose a universal robustification framework for AdvDA that takes a pretrained, already domain-adapted model and fine-tunes it on adversarially transformed inputs, with different choices of source and target transformations yielding different defense variants.

We evaluate the framework under two distinct threat models. For the white-box setting, we employ the projected gradient descent (PGD) attack~\cite{madry2017towards}, which applies bounded $\ell_\infty$ perturbations in the input feature space. We choose PGD because it is the standard attack assumed by all existing defenses for DA 
models~\cite{lo2022exploring,zhu2023srouda,wang2025dart}; however, none of these defenses have been evaluated on malware detection under concept drift. For the black-box setting, we employ MalGuise~\cite{ling2024wolf}, a state-of-the-art binary-level evasion attack, which modifies the malware binary through functionality-preserving control-flow mutations. We instantiate the framework with five defense variants: three derived from DART~\cite{wang2025dart}, a leading robust DA method, using PGD perturbations, and two using MalGuise-generated adversarial binaries. Our evaluation spans longitudinal monthly adaptation windows on Windows malware, assessed at three false-positive-rate operating points.

A central question in our framework is whether adversarial training on the source domain transfers robustness to the target domain, or whether perturbing only the target data suffices. We investigate this by systematically varying the source transformation within each defense family, evaluating each variant against the same attack it was trained to defend. For DART-based defenses, source adversarial training proves essential: variants that perturb both source and target reduce PGD attack success from $28.6\%$ to as low as $3.2\%$. For MalGuise-based defenses, the opposite holds: target-only training already reduces MalGuise attack success to $5.1\%$ with clean true positive rate comparable to the undefended model, while adding source perturbation yields only a marginal improvement ($3.2\%$) at the cost of a $32$--$44\%$ drop in malware detection accuracy and $76.8\times$ training overhead.

Beyond source transferability, we cross-evaluate all defenses against both attacks and find that robustness does not transfer across our evaluated threat models: DART-based defenses offer no protection against MalGuise, and MalGuise-based defenses remain fully vulnerable to PGD. Finally, we assess the practical trade-offs of each defense in terms of clean true positive rate and computational cost, and provide concrete deployment recommendations that balance robustness, performance, and efficiency.

In summary, our contributions are:
\begin{enumerate}
    \item To the best of our knowledge, this is the first adversarial robustness study of a DA-based malware detector, and the first to evaluate both PGD and MalGuise attacks against a malware detector operating under concept drift.
    \item We propose a universal robustification framework for AdvDA that is agnostic to the attack type and the choice of input transformations, and instantiate it with five defense variants across two threat models.
    \item Through extensive experiments (9 defense configurations, 5 adaptation windows, 3 FPR operating points), we reveal two key findings: (i) source adversarial training is essential for DART-based defenses but counterproductive for MalGuise-based defenses, and (ii) robustness does not transfer between the white-box PGD and black-box MalGuise threat models.
\end{enumerate}

\section{Background and Related Work}

\subsection{Drift Adaptation for Windows Malware Detection}
As malware evolves over time, concept drift degrades the accuracy of deployed detectors~\cite{galloro2022systematical}. To maintain detection accuracy, models are periodically retrained on recent samples. The main obstacle is annotation cost: labeling new malware requires either manual reverse-engineering by domain experts or time-intensive dynamic analysis in sandboxed environments, neither of which scales~\cite{botacin2025towards,tripathi2025towards}.

A common strategy to deal with limited labeling budgets is to carefully select which newly arriving samples to annotate, prioritizing those whose labels would be most informative for updating the model. Various selection criteria have been proposed~\cite{chen2023continuous,tripathi2025towards,abusnaina2025exposing,barbero2022transcending,yang2021cade,han2023anomaly,he2025combating}. Once the selected samples are labeled, the model is typically updated through one of two simple procedures: cold-start retraining, which discards the old model and trains a fresh one on all available data, or warm-start fine-tuning, which continues training the existing model on the new labels~\cite{chen2023continuous}. Li et al.~\cite{li2025revisiting} show that neither strategy is effective in label-scarce situations and propose a new training algorithm based on adversarial domain adaptation (AdvDA). AdvDA works by simultaneously minimizing two objectives: (1) classification error on both old and new labeled samples, and (2) discrepancy between the feature distributions of those two domains, making the representations invariant to distributional shift. However, neither AdvDA nor any of the other continual-learning methods consider adversarial robustness. We focus on AdvDA because it is the best-performing drift-adaptive detector under label scarcity, and its explicit source-target alignment raises a unique question about how robustness transfers across its aligned domains.

\subsection{Attacking DNNs for Malware Detection}\label{sec:attack}
Adversarial attacks on malware detectors span multiple dimensions: they may target static or behavioral/dynamic analysis pipelines~\cite{digregorio2024tarallo,d2020dissection}, operate at training time (e.g., backdooring~\cite{d2023lookin}) or test time, and assume varying levels of attacker access. We focus on test-time evasion attacks against static detectors.

Early adversarial attacks on malware detectors operate directly on raw bytes. Kreuk et al.~\cite{kreuk2018adversarial} append or modify bytes in non-executable regions of the binary, while IPR~\cite{lucas2021malware} replaces instructions with semantically equivalent alternatives, and Disp~\cite{lucas2021malware} relocates instruction chunks to new locations, linking them with jump instructions and filling the vacated space with NOPs. Defending against these three attacks is a well-studied problem: existing work has demonstrated that adversarial training can effectively improve the robustness of malware detectors against them~\cite{lucas2023adversarial,lucas2024training}.

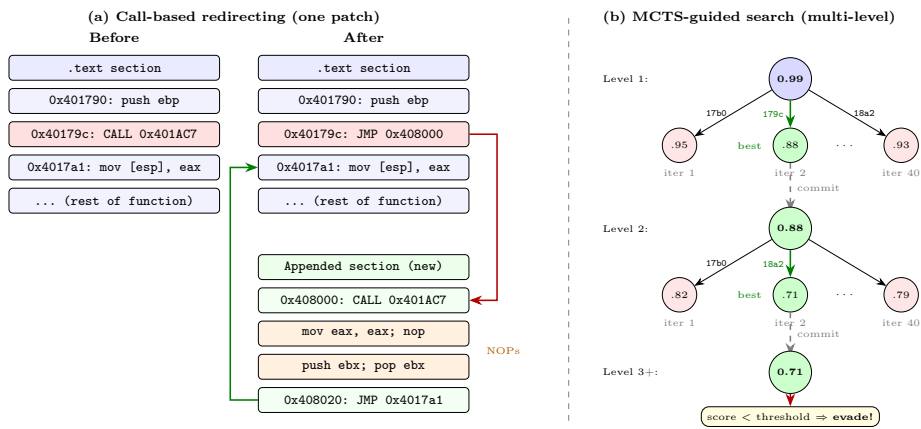
\begin{figure}[t]
\centering
\resizebox{\textwidth}{!}{%
\begin{tikzpicture}[
    >=Stealth,
    box/.style={draw, rounded corners=2pt, minimum width=3.2cm, minimum height=0.45cm, font=\scriptsize\ttfamily, inner sep=2pt},
    phase/.style={font=\scriptsize\bfseries, text=black},
    every node/.style={font=\scriptsize},
    arrow/.style={->, thick},
]

\node[phase, anchor=west] at (-0.3, 6.1) {(a) Call-based redirecting (one patch)};

\node[phase, anchor=south] at (0.3, 5.55) {Before};
\node[box, fill=blue!8, minimum width=3.8cm] (b_title) at (0.3, 5.2) {.text section};
\node[box, fill=blue!5, minimum width=3.8cm] (b_pre)  at (0.3, 4.6) {0x401790: push ebp};
\node[box, fill=red!12, minimum width=3.8cm] (b_call) at (0.3, 4.0) {0x40179c: CALL 0x401AC7};
\node[box, fill=blue!5, minimum width=3.8cm] (b_post) at (0.3, 3.4) {0x4017a1: mov [esp], eax};
\node[box, fill=blue!5, minimum width=3.8cm] (b_more) at (0.3, 2.8) {... (rest of function)};

\node[phase, anchor=south] at (4.8, 5.55) {After};
\node[box, fill=blue!8, minimum width=3.8cm] (a_title) at (4.8, 5.2) {.text section};
\node[box, fill=blue!5, minimum width=3.8cm] (a_pre)  at (4.8, 4.6) {0x401790: push ebp};
\node[box, fill=red!12, minimum width=3.8cm] (a_call) at (4.8, 4.0) {0x40179c: JMP 0x408000};
\node[box, fill=blue!5, minimum width=3.8cm] (a_post) at (4.8, 3.4) {0x4017a1: mov [esp], eax};
\node[box, fill=blue!5, minimum width=3.8cm] (a_more) at (4.8, 2.8) {... (rest of function)};

\node[box, fill=green!8, minimum width=3.8cm] (new_title) at (4.8, 1.6) {Appended section (new)};
\node[box, fill=green!5, minimum width=3.8cm] (ncall) at (4.8, 1.0) {0x408000: CALL 0x401AC7};
\node[box, fill=orange!12, minimum width=3.8cm] (nop1) at (4.8, 0.4) {mov eax, eax; nop};
\node[box, fill=orange!12, minimum width=3.8cm] (nop2) at (4.8, -0.2) {push ebx; pop ebx};
\node[box, fill=green!5, minimum width=3.8cm] (jmpback) at (4.8, -0.8) {0x408020: JMP 0x4017a1};

\draw[arrow, red!70!black, thick] (a_call.east) -- ++(0.5,0) |- (ncall.east);
\draw[arrow, green!50!black, thick] (jmpback.west) -- ++(-0.5,0) |- (a_post.west);

\node[font=\tiny, text=orange!70!black, anchor=west] at (6.9, 0.1) {NOPs};

\node[phase, anchor=west] at (9.0, 6.1) {(b) MCTS-guided search (multi-level)};

\draw[dashed, gray] (8.5, 6.1) -- (8.5, -1.2);

\node[font=\tiny, anchor=west] at (9.0, 5.0) {Level 1:};
\node[draw, circle, fill=blue!15, minimum size=0.55cm, font=\tiny\bfseries] (root) at (12.5, 5.0) {0.99};

\node[draw, circle, fill=red!10, minimum size=0.5cm, font=\tiny] (c1) at (10.5, 3.8) {.95};
\node[draw, circle, fill=green!20, minimum size=0.5cm, font=\tiny] (c2) at (12.5, 3.8) {.88};
\node[font=\tiny] (dots1) at (13.5, 3.8) {$\cdots$};
\node[draw, circle, fill=red!10, minimum size=0.5cm, font=\tiny] (c3) at (14.5, 3.8) {.93};

\node[font=\tiny, text=gray] at (10.5, 3.3) {iter 1};
\node[font=\tiny, text=gray] at (12.5, 3.3) {iter 2};
\node[font=\tiny, text=gray] at (14.5, 3.3) {iter 40};

\draw[arrow, thin] (root) -- (c1) node[midway, left, font=\tiny] {\texttt{17b0}};
\draw[arrow, thick, green!50!black] (root) -- (c2) node[midway, left, font=\tiny] {\texttt{179c}};
\draw[arrow, thin] (root) -- (c3) node[midway, right, font=\tiny] {\texttt{18a2}};

\node[font=\tiny, text=green!50!black, anchor=east] at (12.1, 3.8) {best};

\draw[->, dashed, thick, gray] (c2) -- (12.5, 2.6) node[midway, right, font=\tiny, text=gray] {commit};

\node[font=\tiny, anchor=west] at (9.0, 2.3) {Level 2:};
\node[draw, circle, fill=green!20, minimum size=0.55cm, font=\tiny\bfseries] (r2) at (12.5, 2.3) {0.88};

\node[draw, circle, fill=red!10, minimum size=0.5cm, font=\tiny] (c2a) at (10.5, 1.1) {.82};
\node[draw, circle, fill=green!20, minimum size=0.5cm, font=\tiny] (c2b) at (12.5, 1.1) {.71};
\node[font=\tiny] (dots2) at (13.5, 1.1) {$\cdots$};
\node[draw, circle, fill=red!10, minimum size=0.5cm, font=\tiny] (c2n) at (14.5, 1.1) {.79};

\node[font=\tiny, text=gray] at (10.5, 0.6) {iter 1};
\node[font=\tiny, text=gray] at (12.5, 0.6) {iter 2};
\node[font=\tiny, text=gray] at (14.5, 0.6) {iter 40};

\draw[arrow, thin] (r2) -- (c2a) node[midway, left, font=\tiny] {\texttt{17b0}};
\draw[arrow, thick, green!50!black] (r2) -- (c2b) node[midway, left, font=\tiny] {\texttt{18a2}};
\draw[arrow, thin] (r2) -- (c2n);

\node[font=\tiny, text=green!50!black, anchor=east] at (12.1, 1.1) {best};

\draw[->, dashed, thick, gray] (c2b) -- (12.5, 0.0) node[midway, right, font=\tiny, text=gray] {commit};

\node[font=\tiny, anchor=west] at (9.0, -0.3) {Level 3+:};
\node[draw, circle, fill=green!20, minimum size=0.55cm, font=\tiny\bfseries] (r3) at (12.5, -0.3) {0.71};

\node[draw, rounded corners, fill=yellow!15, minimum width=2.2cm, font=\tiny] (exit) at (12.5, -1.1) {score $<$ threshold $\Rightarrow$ \textbf{evade!}};
\draw[arrow, thick, red!70!black] (r3) -- (exit);

\end{tikzpicture}%
}
\caption{MalGuise attack overview. \textbf{(a)}~A single call-site patch: the original \texttt{CALL} is redirected via a \texttt{JMP} to an appended section containing the preserved call, semantic NOPs, and a \texttt{JMP} back. The example corresponds to the best node selected in Level~1 of~(b). \textbf{(b)}~MCTS searches over which call sites to patch across multiple levels; at each level, 40 iterations explore candidates and the best is committed as the new root.}
\label{fig:malguise}
\end{figure}

As defenses against these byte-level attacks have matured, more recent attacks target the control-flow graph (CFG) of malware binaries, enabling harder-to-detect perturbations. SRL~\cite{zhang2022semantics} modifies CFG nodes by injecting semantic NOPs, while MalGuise~\cite{ling2024wolf} manipulates both nodes and edges, achieving the highest attack success rate among all existing attacks, including SRL, IPR, and Disp. MalGuise uses Monte Carlo Tree Search (MCTS) to find functionality-preserving modifications that cause a classifier to misclassify malware as benign (Figure~\ref{fig:malguise}). Given a malware PE binary, MalGuise extracts the locations of all \texttt{CALL} instructions. It then runs MCTS over two dimensions: \emph{levels} control the number of call sites patched in sequence, and a per-level \emph{budget} of iterations determines how thoroughly each candidate modification is explored. At each iteration, MCTS selects a call site and patches it by redirecting the \texttt{CALL} through a \texttt{JMP} to the appended section, where the original call is preserved, random semantic NOPs are injected, and a \texttt{JMP} returns control to the original flow. The modified binary is then scored by the classifier. After exhausting the budget at a given level, the algorithm commits to the best-scoring modification and advances to the next level, progressively adding patches. The search terminates early if the detector's confidence drops below its threshold at any point.

Although MalGuise demonstrates strong evasion against malware detectors in the typical supervised setting, where models are trained once, it has not been evaluated against adaptive detectors specifically designed to handle concept drift.

\subsection{Attacking and Defending Domain Adaptation Models}
Several methods have been proposed to improve the adversarial robustness of domain adaptation models~\cite{lo2022exploring,zhu2023srouda,wang2025dart}. These methods operate under the unsupervised domain adaptation (DA) setting where the target domain is entirely unlabeled, and therefore rely on pseudo-labels or self-supervision to enable adversarial training. Among them, DART~\cite{wang2025dart} proposes a unified defense framework that can be combined with different unsupervised DA methods, achieving state-of-the-art robustness. In our setting, a small number of labeled target samples are available, which allows us to apply adversarial training directly. As we show in Section~\ref{framework}, all three methods~\cite{lo2022exploring,zhu2023srouda,wang2025dart} reduce to the same generic defense framework under this assumption. Additionally, all existing defenses for DA 
models evaluate robustness against white-box PGD attacks~\cite{madry2017towards}, which iteratively perturb inputs along the gradient of the classification loss to maximize misclassification within an $\ell_\infty$-norm constraint. Moreover, these defenses have only been evaluated on general image classification tasks, not on malware detection under concept drift.

\subsection{Threat Models}
We consider two adversary models that share the same goal: causing a malware classifier to incorrectly classify malware as benign. They differ in the attacker's knowledge of and access to the target system.

\subsubsection{White-Box Attacker (PGD).} The attacker has full access to the model architecture, parameters, and gradients, but cannot modify or influence the trained model in any way except by changing the input to the classifier. We adopt PGD~\cite{madry2017towards} as described above. Note that some prior adversarial attacks on malware detectors~\cite{kreuk2018adversarial,lucas2021malware} also assume white-box access, but their perturbations operate on the malware binary itself through functionality-preserving transformations rather than applying PGD-style perturbations on the input feature space. We consider PGD attacks for two reasons: (1)~all existing defenses for unsupervised DA
models assume a white-box PGD attack~\cite{lo2022exploring,zhu2023srouda,wang2025dart}, and we want to test whether defenses derived from them remain effective when applied to malware detection under concept drift, particularly in our setting where a small number of labeled target samples are available; and (2)~to our knowledge, the adversarial robustness of malware detectors has never been evaluated under PGD attacks, and doing so enables a direct comparison with the binary-level MalGuise attack.

\subsubsection{Black-Box Attacker (MalGuise).} The attacker has no knowledge of the training data, learning algorithm, model architecture, or model weights. The attacker only knows the type of features (e.g., images, graphs) used to represent the executable and can query the deployed detector for a prediction score. To ensure that realistic adversarial malware can be generated, the adversary is restricted to manipulating Windows PE executables while adhering to the PE format specification, so that modified binaries remain functional. We instantiate this threat model with MalGuise~\cite{ling2024wolf}, which serves as a realistic black-box threat for evaluating drift-adaptive malware detectors.

\section{Technical Approach}
We propose a universal robustification framework for AdvDA and instantiate it with five defense variants: three based on DART with PGD perturbations, and two based on adversarial training with MalGuise. We begin by reviewing the AdvDA baseline.

Recall that the original AdvDA~\cite{li2025revisiting}, without any defense, is based on neural networks consisting of three components: a generator $g$, a label classifier $f$, and a domain discriminator $d$. Given source data $(X_S, Y_S)$ and limited labeled target data $(X_T, Y_T)$, the label prediction for a sample $x$ is $(f \circ g)(x)$. The key insight is that if the feature representations produced by $g$ are domain-invariant, the domain divergence will be small. To measure this, AdvDA trains a discriminator $d$ that tries to classify source examples as $\mathbf{0}$ and target examples as $\mathbf{1}$ based on the representations from $g$, and uses its negated loss as an empirical proxy for domain divergence, denoted $\Omega$. Let $\mathcal{L}$ denote the cross-entropy loss. The overall objective is:

\begin{equation}
\min_{g,f} \underbrace{\mathcal{L}(f \!\circ\! g (X_S);\, Y_S)}_{\text{Source Loss}} + \lambda_1 \underbrace{\mathcal{L}(f \!\circ\! g (X_T);\, Y_T)}_{\text{Target Loss}} + \lambda_2 \underbrace{\Omega(X_S,\, X_T,\, g)}_{\text{Domain Divergence}}
\end{equation}

where the empirical proxy for domain divergence is defined as:

\begin{equation}\label{eq:omega}
\Omega(X_S,\, X_T,\, g) = \sup_{d} \left[ -\mathcal{L}\bigl((d \circ g)(X_S),\, \mathbf{0}\bigr) - \mathcal{L}\bigl((d \circ g)(X_T),\, \mathbf{1}\bigr) \right]
\end{equation}

\subsection{Universal Robustification Framework}\label{framework}

We derive our framework from DART~\cite{wang2025dart}, which provides a principled approach to adversarially robust unsupervised DA
and has been shown to outperform other defense methods~\cite{lo2022exploring,zhu2023srouda}. DART 
uses pseudo-labeling to obtain pseudo-labels for unlabeled target training data. In AdvDA, the target training data already has a few labeled samples available. This allows us to derive a new robustification framework without the need for pseudo-labeling, and we further generalize it into a universal robustification framework for AdvDA that is agnostic to (i) the specific adversarial attack for malware detectors and (ii) the specific source and target transformations that are applied.

Our key idea is to replace the clean inputs $X_S$ and $X_T$ with transformed versions $\tilde{X}_S$ and $\tilde{X}_T$, which can be the identity (i.e., the original clean data) or a transformed version of it. The specific transformations differ across defense variants and we study five different variants in this work (Sections~\ref{dart} and~\ref{malguise}). The robustified AdvDA objective then becomes:
\begin{equation}\label{eq:robust-advda}
\min_{g,f} \; \mathcal{L}\bigl(f \circ g(\tilde{X}_S);\, Y_S\bigr) + \lambda_1 \, \mathcal{L}\bigl(f \circ g(\tilde{X}_T);\, Y_T\bigr) + \lambda_2 \, \Omega(\tilde{X}_S,\, \tilde{X}_T,\, g)
\end{equation}
where $\lambda_1$ controls the weight of the target classification loss and $\lambda_2$ controls the weight of the domain divergence term. The framework is \emph{universal} in the sense that different choices of $\tilde{X}_S$ and $\tilde{X}_T$ yield different defense variants, while the training procedure remains the same.

The optimization is solved via alternating updates, as summarized in Algorithm~\ref{alg:robust-advda}: at each iteration, a mini-batch of source and target examples is sampled, transformed to obtain $\tilde{x}_s$ and $\tilde{x}_t$, and used to (i) update the discriminator $d$ to maximize the domain divergence, and (ii) update the generator $g$ and classifier $f$ to minimize the combined objective. The generator and classifier are warm-started from a pretrained (non-robust) AdvDA model, so the robust training fine-tunes an already domain-adapted model.

\begin{algorithm}[t]
\caption{Robust AdvDA Training}
\label{alg:robust-advda}
\KwIn{Source data $(X_S, Y_S)$, target data $(X_T, Y_T)$, pretrained weights for $g$ and $f$, iterations $T$.}
Initialize: Warm-start $g, f$ from pretrained AdvDA; randomly initialize $d$\;
\For{$t = 1, \ldots, T$}{
    Sample mini-batch $(x_s, y_s) \subset (X_S, Y_S)$ and $(x_t, y_t) \subset (X_T, Y_T)$\;
    Compute transformed inputs $\tilde{x}_s$ and $\tilde{x}_t$ (variant-specific; see Sections~\ref{dart},~\ref{malguise})\nllabel{line:transform}\;
    Forward pass: compute features $g(\tilde{x}_s)$, $g(\tilde{x}_t)$ and predictions from $f$ and $d$\;
    \eIf{discriminator step}{
        Update $d$ to maximize the domain divergence\;
    }{
        Update $g, f$ to minimize Eq.~\eqref{eq:robust-advda}\;
    }
}
\Return{$f \circ g$}\;
\end{algorithm}

In the following subsections, we instantiate the framework with five defense variants that differ in how $\tilde{X}_S$ and $\tilde{X}_T$ are constructed. Three variants use DART-based adversarial training with PGD perturbations (Section~\ref{dart}), and two use adversarial training with MalGuise attack on malware binaries (Section~\ref{malguise}). Across both families, all variants apply an adversarial transformation to the target data but vary the source transformation, from leaving it clean to applying different adversarial perturbations. This design allows us to investigate whether robustness learned on the source domain transfers to the target, or whether adversarial training on the target alone achieves a better trade-off between clean detection performance and adversarial robustness. Since PGD perturbations are computed on-the-fly per mini-batch, we describe the DART variants using the mini-batch notation $(x_s, y_s)$ and $(x_t, y_t)$ from Algorithm~\ref{alg:robust-advda}.

\subsection{DART-Based Adversarial Training with PGD}\label{dart}

The three DART-based defense variants directly derive from DART~\cite{wang2025dart}. They differ only in how the source transformation $\tilde{x}_s$ is constructed. All three share the same target transformation $\tilde{x}_t$, described below. The perturbations are generated via PGD.

PGD generates adversarial examples by iteratively perturbing an input to \emph{maximize} a chosen loss function $J$, subject to an $\ell_\infty$ constraint of radius~$\varepsilon$. Starting from a random initialization inside the $\varepsilon$-ball around the clean input $x$, PGD performs $K$ projected gradient ascent steps:
\begin{equation}\label{eq:pgd}
x^{(k+1)}_{\text{adv}} = \Pi_{x,\varepsilon}\!\Big(x^{(k)}_{\text{adv}} + \alpha \cdot \mathrm{sign}\!\big(\nabla_{x}\, J(x^{(k)}_{\text{adv}})\big)\Big),
\end{equation}
where $\alpha$ is the step size and $\Pi_{x,\varepsilon}$ projects back onto the $\varepsilon$-ball. The specific loss $J$ differs across source and target transformations, as detailed below.

\paragraph{Source Transformations.}
Three choices of $\tilde{x}_s$ are considered, yielding three defense variants:

\begin{itemize}
    \item \textbf{DART (clean).} No perturbation: $\tilde{x}_s = x_s$. Robustness is driven entirely by the adversarial target transformation.

    \item \textbf{DART (adv).} PGD maximizes the classification loss: $J = \mathcal{L}\!\big((f \circ g)(\tilde{x}_s);\, y_s\big)$.

    \item \textbf{DART (kl).} PGD maximizes the KL divergence between predictions on the perturbed and clean inputs: $J = \mathrm{KL}\!\big((f \circ g)(\tilde{x}_s)\;\|\;(f \circ g)(x_s)\big)$.
\end{itemize}

\paragraph{Target Transformation (Shared Across All Variants).}
For all three variants, PGD generates $\tilde{x}_t$ by maximizing $J = \lambda_2\, \Omega(\tilde{x}_s,\, \tilde{x}_t,\, g) + \lambda_1\, \mathcal{L}\!\big((f \circ g)(\tilde{x}_t);\, y_t\big)$, where $\tilde{x}_s$ is the (already computed) source transformation and $\Omega$ is the domain divergence proxy (Eq.~\ref{eq:omega}). The first term increases domain divergence (making the discriminator more accurate), while the second term pushes the classifier toward misclassification. 

\subsection{Adversarial Training with MalGuise}\label{malguise}

The MalGuise variants replace clean malware inputs with features extracted from adversarial binaries produced by the MCTS-based MalGuise evasion attack. Because the adversarial binaries are pre-generated, Line~\ref{line:transform} of Algorithm~\ref{alg:robust-advda} is skipped.

\paragraph{Source Transformations.}
Two choices of $\tilde{X}_S$ are considered, yielding two defense variants:
\begin{itemize}
    \item \textbf{MalGuise (clean).} No perturbation: $\tilde{X}_S = X_S$. Robustness is driven entirely by the adversarial target transformation.

    \item \textbf{MalGuise (adv).} The MalGuise attack is executed on the source malware binaries, and the input features extracted from the bypassed samples replace the clean source malware features.
\end{itemize}

\paragraph{Target Transformation (Shared Across Both Variants).}
For both variants, $\tilde{X}_T$ is constructed by executing the MalGuise attack on the target malware binaries and replacing the clean target malware features with those extracted from the bypassed samples.

Unlike the DART variants, where PGD perturbations are computed on-the-fly per mini-batch and do not depend on the detection threshold (PGD maximizes a loss function regardless of the FPR operating point), the MalGuise adversarial binaries must be pre-generated by running the attack against the pretrained (undefended) AdvDA model. Whether a sample bypasses the detector depends on the threshold, so we run the attack at all three FPR operating points and combine the bypassed samples into a single adversarial training set for $\tilde{X}_S$ and $\tilde{X}_T$, yielding a larger and more diverse set than any threshold alone.

\subsection{Fair Comparison Protocol}\label{comparable}

Having defined the five defense variants, we now describe the evaluation protocol used to compare them fairly. We assess each defense along three axes: \emph{attack success rate} (ASR), which measures adversarial robustness; \emph{clean TPR}, which measures detection performance on unperturbed malware; and \emph{computational cost}, which measures the practical overhead of each defense. Together, these metrics provide a well-rounded view of the trade-off between robustness and performance. The formal definitions are given in Section~\ref{sec:evaluation}; here we describe the construction of the comparable evaluation set used for ASR.

Attack success rate (ASR) measures the ratio of correctly detected malware that an adversary can cause to evade detection after perturbation. By restricting to samples that the model already classifies as malware, ASR isolates the effect of the attack itself from the model's detection failures on the malware. The detection threshold is usually calibrated to a false-positive rate (FPR) operating point on a held-out benign set. As a result, each defense model yields a different threshold for the same FPR, and consequently a different set of detected malware. A na\"ive ASR computation attacks only the samples that a given model detects at its own threshold. This makes ASR comparison across models misleading because the set of malware available for attack differs across models.

To obtain a fair comparison, we restrict the attack evaluation to the \emph{common malware set}: the intersection of malware binaries that \emph{all} defense models correctly detect at a given FPR operating point. Concretely, for each target testing set and each FPR threshold, we identify the set of malware samples that every one of the defense models scores above its respective threshold. Only these samples are attacked, and ASR is computed with this shared denominator. Because the intersection is dominated by the model with the lowest clean TPR, the common set is conservative (it contains only the most confidently detected malware) but it guarantees that every model is evaluated on the same malware set.

\section{Evaluation}\label{sec:evaluation} 
This section aims at answering the following research questions: 
\begin{itemize}
  \item \textbf{RQ1 (PGD attack and defense performance)}: How effectively does DART-based adversarial training reduce PGD attack success under concept drift?
  \item \textbf{RQ2 (MalGuise attack and defense performance)}: How effectively does MalGuise-based adversarial training reduce MalGuise attack success under concept drift?
  \item \textbf{RQ3 (Cross-attack robustness transferability)}: Does robustness gained against one attack type transfer to a different attack type?
  \item \textbf{RQ4 (Practical costs)}: What is the impact of each defense algorithm on clean detection accuracy and computational cost?
  \item \textbf{RQ5 (Source robustness transferability)}: Does adversarial training on the source domain transfer robustness to the target, or does adversarial training on the target alone achieve a better trade-off between clean detection performance and adversarial robustness?
\end{itemize}
\noindent RQ5 is addressed throughout RQ1--RQ4 and synthesized in Section~\ref{sec:summary}.
\subsection{Experiment Setup}
  \subsubsection{Dataset.} Our experiments are conducted on MB-24+~\cite{li2025lfreeda}, an extended variant of MB-24~\cite{li2025revisiting}, the dataset used to evaluate AdvDA, that spans nine months of Windows malware (March--December 2024). The malware samples are sourced from the MalwareBazaar daily feed~\cite{malwarebazaar_api} and are deduplicated by SHA-256 hash both within and across months. Monthly family counts range from $81$ to $126$, with $27\%$--$50\%$ of families in each month appearing for the first time relative to the preceding month (Table~\ref{newmb24}). This high rate of new malware families makes MB-24+ representative of real-world concept drift. The benign portion of the MB-24+ consists of ${16{,}000}$ Windows PE files from clean installations of Windows 8, 10, and 11 as well as from commonly used applications~\cite{li2025lfreeda,li2025revisiting}.

\begin{table}[t]
\caption{Monthly statistics of the MB-24+ dataset. March 2024 serves as the initial month. June is omitted to separate the source and target domains.}
\label{newmb24}
\centering
\begin{tabular}{lrrrr}
\toprule
\multicolumn{1}{c}{\textbf{Month}} & \multicolumn{1}{c}{\textbf{Samples}} & \multicolumn{1}{c}{\textbf{Families}} & \multicolumn{1}{c}{\textbf{\begin{tabular}[c]{@{}c@{}}New \\ Families\end{tabular}}} & \multicolumn{1}{c}{\textbf{\begin{tabular}[c]{@{}c@{}}Unseen \\ (\%)\end{tabular}}} \\
\midrule
Mar 2024 & 1{,}505 & 105 & -  & -  \\
Apr 2024 & 1{,}080 &  81 & 22  & 27  \\
May 2024 & 1{,}496 & 100 & 44  & 44  \\
Jul 2024 & 1{,}618 & 126 & 60  & 48  \\
Aug 2024 & 1{,}613 & 114 & 46  & 40  \\
Sep 2024 & 1{,}337 &  92 & 32  & 35  \\
Oct 2024 & 1{,}444 &  97 & 45  & 46  \\
Nov 2024 & 1{,}210 & 108 & 54  & 50  \\
Dec 2024 & 1{,}302 & 105 & 46  & 44  \\
\bottomrule
\end{tabular}
\end{table}

We replicate the data partitioning of AdvDA~\cite{li2025revisiting} and LFreeDA~\cite{li2025lfreeda} to simulate a deployed malware detector that is periodically retrained as new samples arrive. The \emph{source domain} (pre-drift) comprises malware from March--May 2024, split $75/25$ into training and testing sets. The \emph{target domain} (post-drift) covers July--December 2024; June is deliberately omitted to create a clear temporal gap between the two domains. Adaptation proceeds as a monthly update: the model is retrained on the fixed source-domain training set combined with $500$ randomly sampled target-domain instances (both malware and benign) from month~$t$, and then evaluated on month~$t{+}1$, yielding five evaluation windows (July$\rightarrow$August through November$\rightarrow$December). Target data does not accumulate across windows; each update uses only the current month's target samples alongside the unchanged source data. We choose a budget of $500$ target labels because with fewer samples the learned threshold approaches $1$ at a low FPR, effectively rejecting most inputs and rendering the detector unusable at low false-positive operating points.

Because benign PE files have no collection timestamps, we treat the benign distribution as temporally stable, consistent with~\cite{li2025lfreeda,li2025revisiting}. Of the $16{,}000$ benign samples, half are assigned to the source domain and the other half is equally split between target training and target testing. This yields a malware-to-benign ratio of approximately $0.5{:}1$ in the source domain and $0.3{:}1$ in the target sets, mitigating spatial bias~\cite{pendlebury2019tesseract} by keeping the target training ratio close to the target testing distribution (Table~\ref{mb24_ratio} in the Appendix).

  \subsubsection{Algorithms.} We study 9 defense algorithms.
  \begin{itemize}
    \item \textbf{AdvDA}. This is standard AdvDA without any defense mechanism.
    \item \textbf{DART-based}. We experiment with DART-based defense for three different source choices as described in Section~\ref{dart}; {\bfseries DART (clean)}, {\bfseries DART (adv)}, {\bfseries DART (kl)}. We train each DART-based model with two perturbation sizes, resulting in 6 DART-based models.
    \item \textbf{MalGuise-based}. We experiment with adversarial training with MCTS-based MalGuise attack on the binary for two source choices as described in Section~\ref{malguise}; namely, \textbf{MalGuise (clean), MalGuise (adv)}. 
  \end{itemize}
  
  \subsubsection{Architecture and optimization.} All models share the same backbone: the generator $g$ is an ImageNet-pretrained ResNet-18 that extracts a $512$-dimensional feature vector from each input image. The classifier $f$ is a two-layer fully connected network ($512 \rightarrow 256 \rightarrow 2$) with ReLU and dropout ($p{=}0.5$), while the domain discriminator $d$ uses a wider hidden layer ($512 \rightarrow 1024 \rightarrow 2$) with batch normalization and ReLU. We train all models for $30$ epochs using cross-entropy loss, the Adam optimizer (learning rate $10^{-4}$), a batch size of $32$, domain divergence weight $\lambda_2{=}0.02$, target weight $\lambda_1{=}0.5$. These hyperparameters were selected from $10$ configurations. Tuning and configuration details are provided in Table~\ref{tab:hyperparam} in the Appendix. The DART-based and MalGuise-based adversarial training use the same hyperparameters. Our DART implementation is based on the DomainRobust codebase~\cite{domainrobust}.

 \subsubsection{Attack setup.} For the PGD attack, we assume an $\ell_\infty$-norm perturbation set $\mathcal{B}(x) = \{\tilde{x} : \|\tilde{x} - x\|_\infty \leq \varepsilon\}$ and experiment with two values of $\varepsilon$: we report results for $\varepsilon = 2/255$ and for $\varepsilon = 8/255$. During training, adversarial examples are generated using $5$ steps of PGD with a random start within the $\varepsilon$-ball and a step size of $\varepsilon/4$ (i.e., $1/255$ for $\varepsilon{=}2/255$ and $2/255$ for $\varepsilon{=}8/255$), ensuring the perturbation boundary is reachable within the allotted steps. At evaluation time, we \textbf{cross-evaluate} models trained under one PGD strength against the other: specifically, we test whether DART ($\varepsilon{=}2/255$) models hold up against the stronger $\varepsilon{=}8/255$ attack, and how DART ($\varepsilon{=}8/255$) models perform under the weaker $\varepsilon{=}2/255$ attack. This cross-budget evaluation simulates an adaptive attacker who deploys a stronger perturbation budget than the one anticipated during training. All evaluation attacks use $20$ PGD iterations, $4\times$ more than during training, to further ensure the evaluation attack is strictly stronger than the one seen during training.  All attacks have full access to the model parameters (white-box setting). All PGD hyperparameters are chosen to follow~\cite{wang2025dart}.

  For the MCTS-based MalGuise attack~\cite{ling2024wolf}, we target 32-bit PE binaries in a black-box setting: the attack queries the model and receives only the malware confidence score. Because our classifier operates on image representations, each modified binary is converted to its corresponding image before being scored by the model via local inference. Following the default configuration of the original MalGuise, we use $6$ MCTS levels with a budget of $40$ iterations per level, a single simulation per expansion, and a maximum file size increase of $5\%$. The search terminates early if the model's score drops below the detection threshold at any level. We parallelize the attack across $30$ workers to reduce execution time. Even so, attacking with $40$ iterations per level already incurs considerable runtime (Section~\ref{cost}), making substantially larger MCTS budgets impractical for a real-world attacker.

 \subsubsection{Evaluation Metrics.}
  We report ASR and Clean TPR at three FPR operating points: $0.5\%$, $1\%$, and $2\%$, as they are typical operating points for malware detectors. For each defense model, the detection threshold at each FPR is calibrated on the benign samples from the source-domain test set, which is shared across all models.
  \begin{itemize}
    \item \textbf{Attack success rate:} the fraction of samples in the common malware set (defined in Section~\ref{comparable}) that evade detection after attack at the given FPR threshold. Table~\ref{tab:common_set} reports the common malware set size for each FPR and target-test month.
    \item \textbf{Clean TPR:} the true-positive rate on the entire clean target test malware. This is crucial as strong robustness usually hurts clean TPR.
    \item \textbf{Computational cost:} the total time to train each defense model, including adversarial sample generation. This is the computational overhead of each defense algorithm.
  \end{itemize}

  \begin{table}[t]
    \caption{Number of common malware samples across all 9 defense models, by FPR and testing month.}
    \label{tab:common_set}
    \centering
    \small
    \begin{tabular}{l*{5}{ccc}}
    \toprule
    & \multicolumn{3}{c}{\textbf{Aug}} & \multicolumn{3}{c}{\textbf{Sep}} & \multicolumn{3}{c}{\textbf{Oct}} & \multicolumn{3}{c}{\textbf{Nov}} & \multicolumn{3}{c}{\textbf{Dec}} \\
    \cmidrule(lr){2-4} \cmidrule(lr){5-7} \cmidrule(lr){8-10} \cmidrule(lr){11-13} \cmidrule(lr){14-16}
    \textbf{FPR} & 0.5\% & 1\% & 2\% & 0.5\% & 1\% & 2\% & 0.5\% & 1\% & 2\% & 0.5\% & 1\% & 2\% & 0.5\% & 1\% & 2\% \\
    \midrule
    & 97 & 270 & 453 & 54 & 188 & 317 & 85 & 332 & 482 & 257 & 331 & 541 & 130 & 224 & 359 \\
    \bottomrule
    \end{tabular}
\end{table}

\subsection{RQ1: PGD Attack and Defense Performance}
We evaluate AdvDA (no defense) and 6 DART variants: DART (clean), DART (adv), and DART (kl), each trained at two perturbation budgets ($\varepsilon{=}2/255$ and $\varepsilon{=}8/255$), across the five adaptation windows. In total, this yields $7 \times 5 = 35$ model instances, each attacked with PGD at both $\varepsilon{=}2/255$ and $\varepsilon{=}8/255$ for 20 iterations on the common malware set of its corresponding target-test month. We apply each model's detection threshold at the corresponding FPR to classify each adversarial sample as bypassed or detected. The ASR results are presented in Figure~\ref{fig:asr_fpr_pgd_eps2} and Figure~\ref{fig:asr_fpr_pgd_eps8}. We observe that:
\begin{figure}[t]
  \centering
  \begin{subfigure}[b]{\linewidth}
    \centering
    \includegraphics[width=\linewidth]{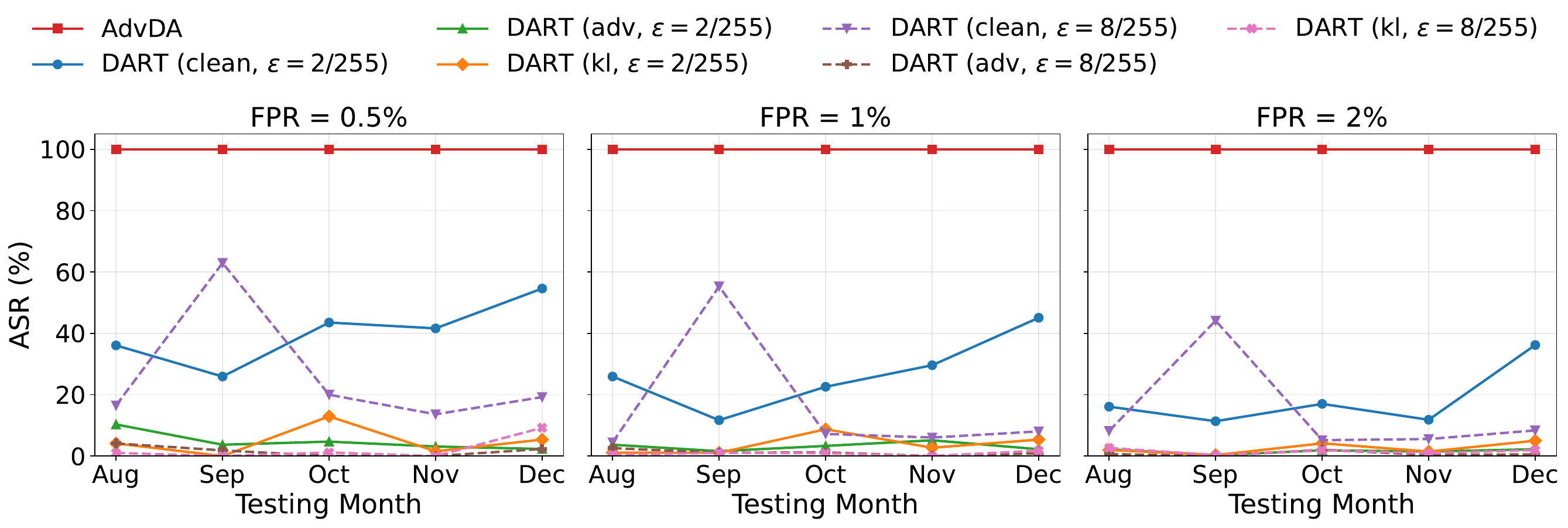}
    \caption{PGD attack, $\varepsilon = 2/255$}
    \label{fig:asr_fpr_pgd_eps2}
  \end{subfigure}
  \vspace{0.5em}
  \begin{subfigure}[b]{\linewidth}
    \centering
    \includegraphics[width=\linewidth]{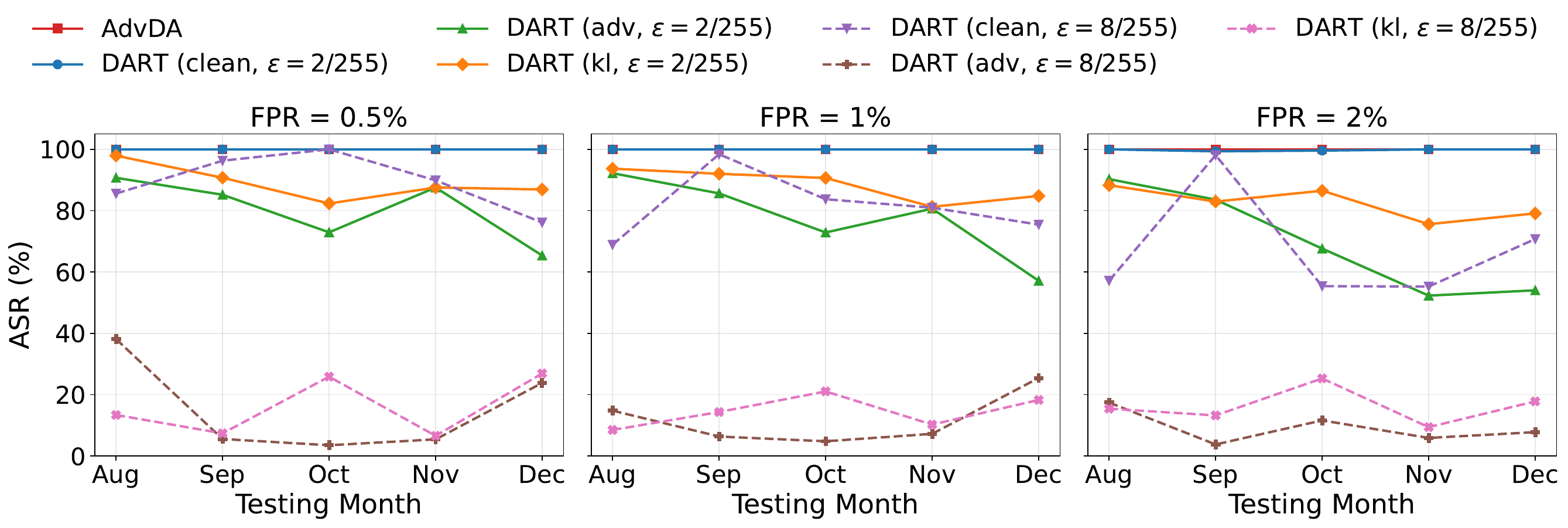}
    \caption{PGD attack, $\varepsilon = 8/255$}
    \label{fig:asr_fpr_pgd_eps8}
  \end{subfigure}
  \caption{ASR of AdvDA and DART-based models at three FPR operating points under PGD attack.}
  \label{fig:asr_fpr_pgd}
\end{figure}

\begin{itemize}
  \item The AdvDA model is bypassed $100\%$ across all testing months and at both PGD attack strengths, confirming that the baseline detector without adversarial training is completely vulnerable to the PGD attack.
  \item Under the weaker PGD attack ($\varepsilon{=}2/255$), all DART variants substantially reduce ASR. Even the least robust variant, DART (clean) trained at $\varepsilon{=}2/255$, still incurs an average ASR of 40\% at 0.5\% FPR, 27\% at 1\% FPR, and 18\% at 2\% FPR, while DART (adv) and DART (kl) reduce ASR to 2--5\% on average across operating points. Models trained at the higher perturbation budget ($\varepsilon{=}8/255$) generally yield even lower ASR under this weaker attack, as expected from exposure to stronger perturbations during training.
  \item Under the stronger PGD attack ($\varepsilon{=}8/255$), models trained at the lower budget ($\varepsilon{=}2/255$) degrade substantially: DART (clean) reaches $100\%$ ASR, and even DART (adv) averages $70\%$ ASR at 2\% FPR. In contrast, models trained at the matching budget ($\varepsilon{=}8/255$) remain far more robust, with DART (adv) averaging $9\%$ ASR at 2\% FPR. The sharp degradation reflects robustness overfitting: the model learns to defend within a specific perturbation radius but fails to generalize beyond it. This suggests that robustness in drift-adaptive settings is highly local in perturbation space, and training with stronger perturbations is necessary under stronger attackers.
  \item Across both attack strengths, the robustness ranking among DART variants is consistent: DART (adv) $>$ DART (kl) $>$ DART (clean). DART (adv) generates source perturbations by maximizing the classification loss, directly approximating the PGD attack objective at test time. DART (kl) instead maximizes KL divergence between clean and perturbed predictions, which regularizes the decision boundary but does not explicitly target the classification loss exploited by PGD. DART (clean), lacking any source-side adversarial pressure, relies on target perturbations, yielding the weakest robustness.
\end{itemize}

\subsection{RQ2: MalGuise Attack and Defense Performance}\label{sec:rq2}
We evaluate AdvDA (no defense), MalGuise (clean), and MalGuise (adv) across the five adaptation windows. In total, this yields $3 \times 5 = 15$ model instances, each attacked with the MCTS-based MalGuise attack in a black-box setting on the common malware set of its corresponding target-test month. We apply each model's detection threshold at the corresponding FPR to classify each adversarial sample as bypassed or detected. The ASR results are presented in Figure~\ref{fig:asr_fpr_mcts}. We observe that:

\begin{figure}[t]
  \centering
  \includegraphics[width=\linewidth]{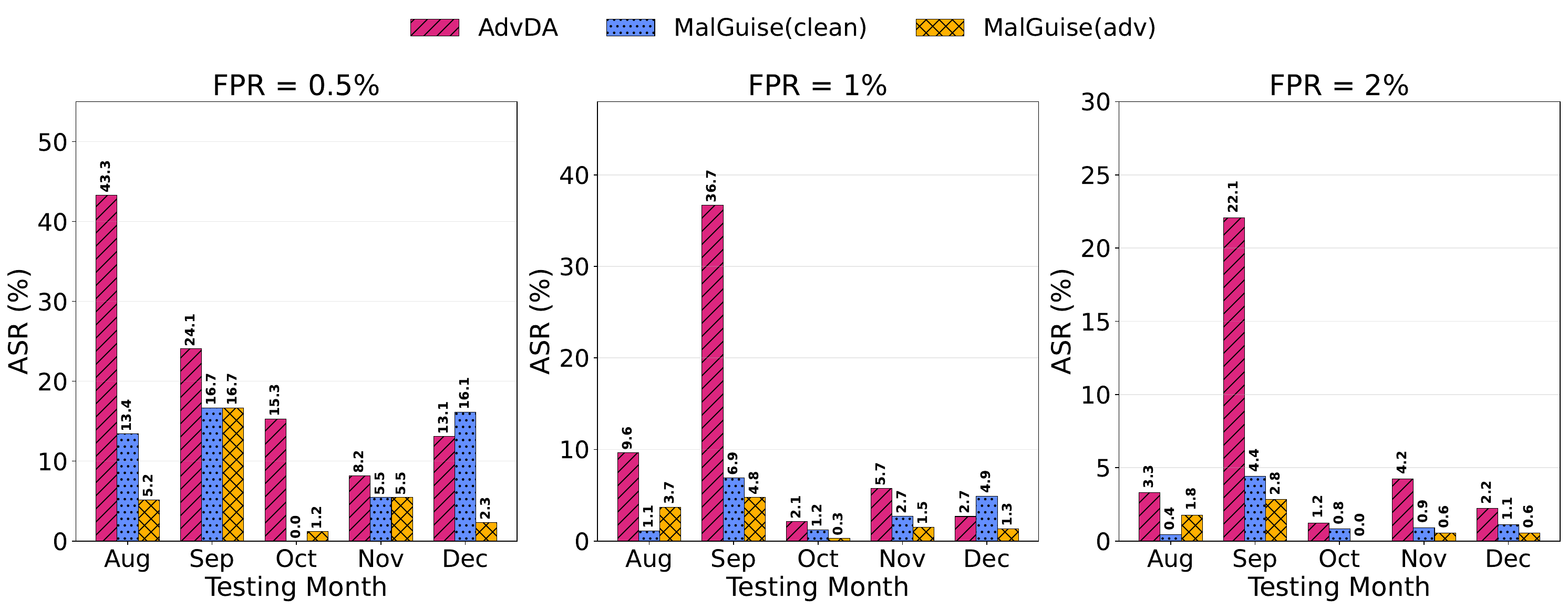}
  \caption{ASR of AdvDA and MalGuise-based models at three FPR operating points under MCTS-based MalGuise attack.}
  \label{fig:asr_fpr_mcts}
\end{figure}
\begin{itemize}
  \item The relatively low ASR of MalGuise against AdvDA ($13\%$ on average across all testing months and FPR operating points)\footnote{We obtained the official MalGuise implementation from the authors~\cite{ling2024wolf}. As a sanity check, we reproduced their reported ASR on their provided MalConv model and threshold value at $1\%$ FPR using our common malware test set, achieving over $90\%$ ASR, consistent with their results. Note that MalConv is a 1D-CNN operating directly on raw bytes, without a ResNet backbone or drift adaptation.} suggests that the model's architecture acts as an implicit defense. Unlike PGD, which perturbs all input dimensions, MalGuise introduces localized structural changes in the binary that translate into spatially sparse perturbations in the image representation. These are largely attenuated by (i) the ResNet-18 backbone, which, through ImageNet pretraining, has learned to capture global structural patterns and is inherently insensitive to sparse, localized perturbations, and (ii) the generator $g$, which further compresses input variations into domain-invariant features. As a result, the attack signal available to guide MCTS is weak, limiting its effectiveness.
  \item Both MalGuise variants substantially reduce ASR compared to AdvDA: MalGuise (adv) to $3\%$ and MalGuise (clean) to $5\%$ on average, with MalGuise (adv) attaining the lowest ASR in 12 out of 15 month-FPR combinations. However, the limited additional gain from source adversarial training reflects a key difference from the PGD setting. PGD perturbations are input-agnostic and transferable, so robustness learned on source inputs generalizes to the target domain. MalGuise attacks, by contrast, are binary-specific rather than domain-general, so adding source adversarial training contributes little additional coverage. As we show in Section~\ref{sec:summary}, this marginal gain comes at a significant cost in clean TPR and computational overhead.
\end{itemize}



\subsection{RQ3: Cross-Attack Robustness Transferability}\label{cross}

RQ1 showed that DART-based adversarial training effectively defends against PGD, and RQ2 showed that MalGuise-based adversarial training reduces MalGuise attack success. An intriguing question is whether robustness gained against one attack type transfers to the other. We investigate this in two directions. First, we attack MalGuise (clean) and MalGuise (adv) with PGD at both $\varepsilon{=}2/255$ and $\varepsilon{=}8/255$ across the five adaptation windows ($2 \times 5 = 10$ model instances). Second, we attack all six DART variants with the MCTS-based MalGuise attack ($6 \times 5 = 30$ model instances). Both evaluations are conducted on the common malware set of each corresponding target-test month. Figure~\ref{fig:rq3_cross} reports the ASR averaged across the five adaptation windows for all nine defense models: rows correspond to the three attack settings (PGD $\varepsilon{=}2/255$, PGD $\varepsilon{=}8/255$, MalGuise) and columns to the three FPR operating points. We observe that:

\begin{figure}[t]
  \centering
  \includegraphics[width=\linewidth]{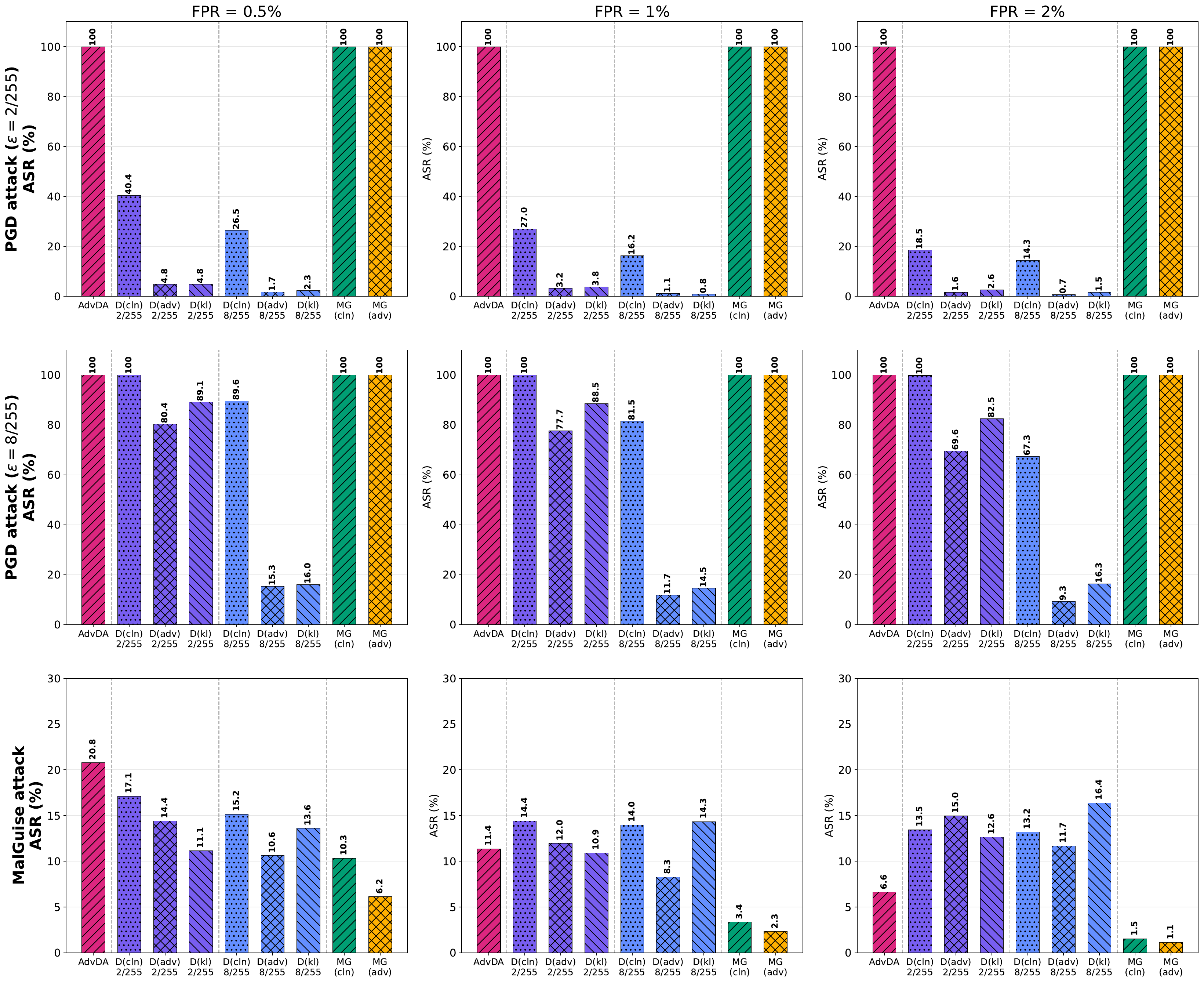}
  \caption{Cross-attack ASR (averaged across five adaptation windows) for all nine defense models.}
  \label{fig:rq3_cross}
\end{figure}

\begin{itemize}
  \item   MalGuise-based adversarial training provides no robustness against PGD. Both MalGuise (clean) and MalGuise (adv) are bypassed $100\%$ under PGD at both $\varepsilon{=}2/255$ and $\varepsilon{=}8/255$, identical to the undefended AdvDA. 
  \item DART-based adversarial training does not consistently improve robustness against the MalGuise attack. At 1\% and 2\% FPR, nearly all DART variants have ASR comparable to or higher than AdvDA. Training with a larger perturbation budget ($\varepsilon{=}8/255$) does not reliably outperform the smaller budget ($\varepsilon{=}2/255$) against the MalGuise attack.
\end{itemize}

The lack of cross-attack robustness reveals a fundamental limitation of adversarial training in drift-adaptive systems: robustness is not a property of the model alone, but of the model-threat pairing. PGD and MalGuise induce fundamentally different perturbation geometries: dense, norm-bounded perturbations across every pixel versus sparse, structure-preserving binary transformations. Adversarial training against one perturbation family does not generalize to the other.


\subsection{RQ4: Practical Costs}
\subsubsection{Clean TPR}
\begin{figure}[t]
  \centering
  \includegraphics[width=0.8\linewidth]{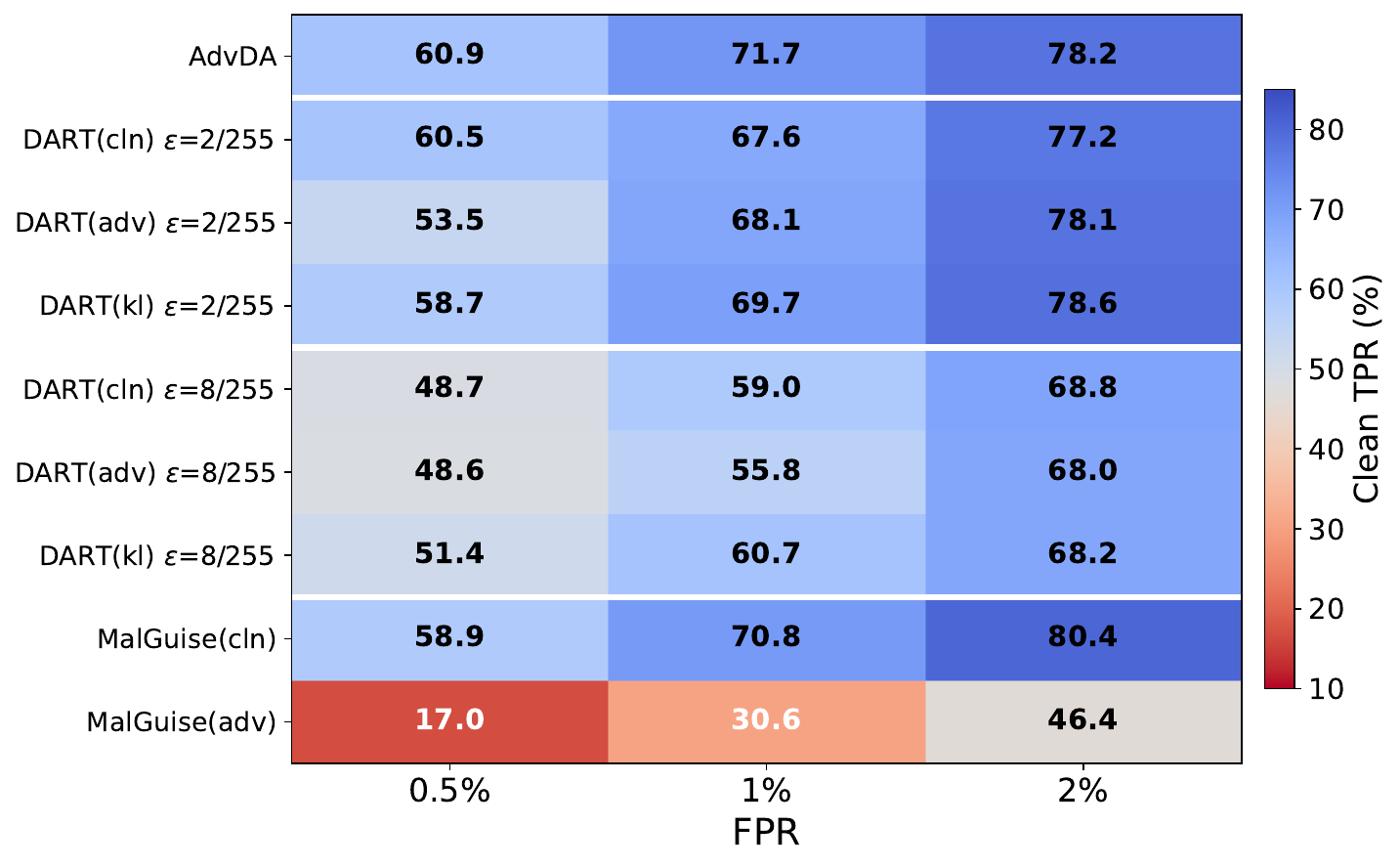}
  \caption{Average clean TPR (\%) across five adaptation windows for all nine defense models. Higher is better. Red indicates low TPR; blue indicates high TPR.}
  \label{fig:rq4_tpr}
\end{figure}

Figure~\ref{fig:rq4_tpr} reports the average clean TPR across the five adaptation windows for all nine defense models at each FPR operating point.

DART ($\varepsilon{=}2/255$) variants maintain TPR comparable to AdvDA ($\Delta \leq 8$\%), with DART (kl) achieving the highest TPR at 0.5\% and 1\% FPR, and DART (kl) and DART (adv) tied at 2\% FPR. DART ($\varepsilon{=}8/255$) variants incur a moderate TPR drop ($\Delta = 10$--$16$\%), with DART (kl) highest at 0.5\% and 1\% FPR, and DART (clean) highest at 2\% FPR. Among MalGuise variants, MalGuise (clean) preserves TPR comparable to AdvDA ($\Delta \leq 4$\%), while MalGuise (adv) suffers a severe TPR reduction ($\Delta = 32$--$44$\%), which is the primary practical cost of this defense. The TPR degradation under stronger adversarial training reflects the classic robustness--accuracy tradeoff, amplified in the drift-adaptation setting: adversarial training reduces sensitivity to perturbations but also limiting the model's ability to distinguish malware from benign samples. For MalGuise (adv), the effect is more severe because source adversarial training replaces clean source features with features from attacked binaries that, as discussed in Section~\ref{sec:rq2}, are binary-specific and do not generalize. The model therefore trains on  source representations that are representative of neither clean nor target-domain inputs, degrading its ability to classify clean samples.


\subsubsection{Computational Cost}\label{cost}
Table~\ref{tab:training_time} reports the average total cost per adaptation window for each defense model. All times are measured on an NVIDIA RTX 3090. DART variants are reported using the $\varepsilon{=}2/255$ configuration; the $\varepsilon{=}8/255$ configuration yields similar times, as the two settings differ only in scalar multipliers within the PGD update and share the same number of attack iterations, architecture, and batch size. Note that the DART training time already includes the cost of generating PGD adversarial samples, since perturbations are computed on-the-fly within each training step rather than in a separate preprocessing stage. Within the DART family, training time increases from DART (clean) to DART (adv) to DART (kl) because each variant performs progressively more PGD computation per step: DART (clean) perturbs only target inputs, DART (adv) perturbs both source and target, and DART (kl) additionally computes a KL-divergence term over the clean and adversarial predictions.

\begin{table}[t]
\caption{Average total cost (hours) per adaptation window for each defense model. }
\label{tab:training_time}
\centering
\resizebox{\linewidth}{!}{%
\begin{tabular}{lccccc}
\toprule
\textbf{Defense model} & \textbf{Model training (h)} & \textbf{MG atk source (h)} & \textbf{MG atk target (h)} & \textbf{Total cost (h)} & \textbf{Slowdown} \\
\midrule
AdvDA                           & 0.24 & --   & --   & 0.24  & $\mathbf{1\times}$ \\
\midrule
DART (clean) $\varepsilon{=}2/255$   & 0.71 & --   & --   & 0.71  & $\mathbf{3.0\times}$ \\
DART (adv) $\varepsilon{=}2/255$     & 0.90 & --   & --   & 0.90  & $\mathbf{3.8\times}$ \\
DART (kl) $\varepsilon{=}2/255$      & 1.08 & --   & --   & 1.08  & $\mathbf{4.5\times}$ \\
\midrule
MalGuise (clean)                 & 0.25 & --   & 0.51 & 0.76  & $\mathbf{3.2\times}$ \\
MalGuise (adv)                   & 0.22 & 17.66 & 0.51 & 18.40 & $\mathbf{76.8\times}$ \\
\bottomrule
\end{tabular}%
}
\end{table}

For MalGuise variants, the total time per adaptation window includes both model training and adversarial sample generation via the MalGuise attack, which must be run prior to training to produce the adversarial training data. As described in Section~\ref{malguise}, the attack is run at all three FPR thresholds to maximize the pool of adversarial samples available for training; all attack times are reported with 30 parallel workers. MalGuise (clean) uses clean source samples and only attacks the target training samples, adding a modest overhead (0.76 h total). MalGuise (adv) additionally attacks the source training samples, making it by far the most expensive defense at 18.40 h per window, of which 17.66 h is spent on source-train adversarial sample generation alone.

\subsection{Summary}\label{sec:summary}

Table~\ref{tab:summary} consolidates the key findings across all five research questions.

\begin{table}[t]
\caption{Summary of defense trade-offs. ASR values are averaged across five adaptation windows and three FPR operating points on the common malware set. Clean TPR and cost are averaged across five windows. $\pm$ denotes standard deviation across windows. $\downarrow$ = lower is better; $\uparrow$ = higher is better.}
\label{tab:summary}
\centering
\resizebox{\linewidth}{!}{%
\begin{tabular}{lccccc}
\toprule
 & \multicolumn{2}{c}{\textbf{PGD ASR $\downarrow$ (\%)}} & & & \\
\cmidrule{2-3}
\textbf{Defense} & $\varepsilon{=}2/255$ & $\varepsilon{=}8/255$ & \textbf{MalGuise ASR $\downarrow$ (\%)} & \textbf{Clean TPR $\uparrow$ (\%)} & \textbf{Cost $\downarrow$ (h)} \\
\midrule
AdvDA (no defense)                       & 100{\scriptsize$\pm$0.0}   & 100{\scriptsize$\pm$0.0}   & 12.9{\scriptsize$\pm$8.8} & 70.3{\scriptsize$\pm$4.7} & 0.24{\scriptsize$\pm$0.01} \\
\midrule
DART (clean) $\varepsilon{=}2/255$           & 28.6{\scriptsize$\pm$9.4}  & 99.9{\scriptsize$\pm$0.1}  & 15.0{\scriptsize$\pm$13.5} & 68.4{\scriptsize$\pm$8.9} & 0.71{\scriptsize$\pm$0.02} \\
DART (adv) $\varepsilon{=}2/255$             & \textbf{3.2}{\scriptsize$\pm$1.2}   & 75.9{\scriptsize$\pm$11.2}  & 13.8{\scriptsize$\pm$13.4} & 66.6{\scriptsize$\pm$4.4} & 0.90{\scriptsize$\pm$0.02} \\
DART (kl) $\varepsilon{=}2/255$              & \textbf{3.7}{\scriptsize$\pm$2.9}   & 86.7{\scriptsize$\pm$4.1}  & 11.6{\scriptsize$\pm$8.8} & 69.0{\scriptsize$\pm$5.3} & 1.08{\scriptsize$\pm$0.02} \\
\midrule
DART (clean) $\varepsilon{=}8/255$           & 19.0{\scriptsize$\pm$17.6}  & 79.5{\scriptsize$\pm$9.5}  & 14.1{\scriptsize$\pm$11.0} & 58.8{\scriptsize$\pm$4.5} & 0.71{\scriptsize$\pm$0.02} \\
DART (adv) $\varepsilon{=}8/255$             & \textbf{1.2}{\scriptsize$\pm$0.7}   & \textbf{12.1}{\scriptsize$\pm$7.6}  & 10.2{\scriptsize$\pm$6.9} & 57.5{\scriptsize$\pm$7.7} & 0.90{\scriptsize$\pm$0.02} \\
DART (kl) $\varepsilon{=}8/255$              & \textbf{1.6}{\scriptsize$\pm$1.5}   & \textbf{15.6}{\scriptsize$\pm$5.9}  & 14.8{\scriptsize$\pm$12.5} & 60.1{\scriptsize$\pm$7.7} & 1.08{\scriptsize$\pm$0.02} \\
\midrule
MalGuise (clean)                         & 100{\scriptsize$\pm$0.0}   & 100{\scriptsize$\pm$0.0}   & \textbf{5.1}{\scriptsize$\pm$3.1}  & 70.0{\scriptsize$\pm$4.2} & 0.76{\scriptsize$\pm$0.02} \\
MalGuise (adv)                           & 100{\scriptsize$\pm$0.0}   & 100{\scriptsize$\pm$0.0}   & \textbf{3.2}{\scriptsize$\pm$2.7}  & 31.3{\scriptsize$\pm$6.3} & 18.40{\scriptsize$\pm$1.96} \\
\bottomrule
\end{tabular}%
}
\end{table}

\paragraph{Cross-Attack Robustness.} DART is the only effective defense against PGD, and MalGuise is the only effective defense against the MalGuise attack. Neither transfers its robustness to the other threat model: MalGuise variants are completely bypassed by PGD, and DART variants offer no improvement over AdvDA under the MalGuise attack.

\paragraph{Source Robustness Transferability.} The benefit of source adversarial training is defense-dependent. For DART, it is essential: DART (clean) incurs 28.6\% ASR under PGD $\varepsilon{=}2/255$, whereas DART (adv) and DART (kl) reduce this to 3.2\% and 3.7\% with comparable TPR. For MalGuise, it is counterproductive: MalGuise (clean) already reduces ASR to 5.1\% with TPR comparable to AdvDA, while MalGuise (adv) gains only a marginal improvement at severe TPR and cost penalties.

\paragraph{Final Recommendations.}
\begin{enumerate}[leftmargin=*, label=(\arabic*)]
  \item \emph{White-box PGD defense.} DART (kl) is preferable to DART (adv): comparable robustness, higher clean TPR, and similar cost. Source adversarial training is essential, and the perturbation budget must match the anticipated attack strength. 
  \item \emph{Black-box MalGuise defense.} MalGuise (clean) offers the best trade-off: 5.1\% ASR with TPR comparable to AdvDA at only $3.2\times$ the training cost. Adding source perturbation incurs a 32--44\% TPR reduction and $76.8\times$ overhead for marginal gain.
\end{enumerate}

\smallskip\noindent\fbox{\parbox{\dimexpr\linewidth-2\fboxsep-2\fboxrule\relax}{\small\textbf{Key takeaway:} No single defense robustifies adaptive malware detectors against both threat models under concept drift. DART and MalGuise address orthogonal threat models, and the role of source adversarial training is defense-dependent: essential for DART, but unnecessary for MalGuise. Practitioners must select their defense according to the anticipated attack vector.}}

\section{Future Work}

Our findings show that no single defense achieves robustness across both gradient-based and structure-preserving attacks, and that robustness is a property of the model--threat pairing rather than the model alone. This motivates a multi-view defense architecture in which specialized detectors, each trained for a different perturbation family, are combined at inference time.

Concretely, one model can be adversarially trained using PGD to defend against feature-space attacks, while another is trained on MalGuise-generated binaries to counter binary-level evasion. At inference time, the ensemble flags a sample as malware if \emph{any} individual model detects it, which is particularly well suited to adversarial settings: it forces the attacker to simultaneously evade all models, which operate in structurally distinct perturbation spaces.

Such an architecture raises the bar for adversaries considerably, as an adaptive attacker would need to jointly optimize across both perturbation spaces, making joint evasion a non-trivial problem. Nevertheless, the additional computational cost and potential effect on FPR must be carefully evaluated. We consider the design and empirical evaluation of multi-view defenses for drift-adaptive malware detectors a promising direction for future work.

\section{Conclusion}

This paper presents the first study of adversarial robustness for drift-adaptive malware detectors, addressing the combined challenge of concept drift and adversarial evasion that prior work has only tackled in isolation. We propose a universal robustification framework that is agnostic to both attack type and input transformation, and instantiate it with five defense variants spanning white-box (PGD) and black-box (MalGuise) threat models. Through evaluation over nine defense configurations, five monthly adaptation windows, and three operating points on MB-24+, we uncover key findings that reshape how practitioners should defend these systems. Robustifying a drift-adaptive detector introduces challenges absent in stationary settings. Source adversarial training is essential for feature-space (PGD) defenses but yields only marginal robustness gains for binary-level (MalGuise) ones at severe cost to clean detection and training efficiency. Notably, the undefended AdvDA already exhibits surprising robustness to the black-box MalGuise attack, which we attribute to the domain-adapted feature representation attenuating localized binary modifications. More broadly, adversarial training designed for attack models from the vision domain does not carry over to malware, underscoring the need for dedicated defenses when robustifying drift-adaptive malware detectors.


\begin{credits}
\subsubsection{\ackname} The work reported in this paper has been supported by the National Science Foundation (NSF) under Grants 2229876 and 2112471.
\end{credits}

%
%
%
\bibliographystyle{splncs04}
\bibliography{references}

@inproceedings{ma2021comprehensive,
  title={A comprehensive study on learning-based PE malware family classification methods},
  author={Ma, Yixuan and Liu, Shuang and Jiang, Jiajun and Chen, Guanhong and Li, Keqiu},
  booktitle={Proceedings of the 29th ACM Joint Meeting on European Software Engineering Conference and Symposium on the Foundations of Software Engineering},
  pages={1314--1325},
  year={2021}
}

@inproceedings{yang2021cade,
  title={$\{$CADE$\}$: Detecting and explaining concept drift samples for security applications},
  author={Yang, Limin and Guo, Wenbo and Hao, Qingying and Ciptadi, Arridhana and Ahmadzadeh, Ali and Xing, Xinyu and Wang, Gang},
  booktitle={30th USENIX Security Symposium (USENIX Security 21)},
  pages={2327--2344},
  year={2021}
}

@inproceedings{barbero2022transcending,
  title={Transcending transcend: Revisiting malware classification in the presence of concept drift},
  author={Barbero, Federico and Pendlebury, Feargus and Pierazzi, Fabio and Cavallaro, Lorenzo},
  booktitle={2022 IEEE Symposium on Security and Privacy (SP)},
  pages={805--823},
  year={2022},
  organization={IEEE}
}

@inproceedings{jordaney2017transcend,
  title={Transcend: Detecting concept drift in malware classification models},
  author={Jordaney, Roberto and Sharad, Kumar and Dash, Santanu K and Wang, Zhi and Papini, Davide and Nouretdinov, Ilia and Cavallaro, Lorenzo},
  booktitle={26th USENIX security symposium (USENIX security 17)},
  pages={625--642},
  year={2017}
}

@inproceedings{han2023anomaly,
  title={Anomaly Detection in the Open World: Normality Shift Detection, Explanation, and Adaptation.},
  author={Han, Dongqi and Wang, Zhiliang and Chen, Wenqi and Wang, Kai and Yu, Rui and Wang, Su and Zhang, Han and Wang, Zhihua and Jin, Minghui and Yang, Jiahai and others},
  booktitle={NDSS},
  year={2023}
}

@inproceedings{pendlebury2019tesseract,
  title={$\{$TESSERACT$\}$: Eliminating experimental bias in malware classification across space and time},
  author={Pendlebury, Feargus and Pierazzi, Fabio and Jordaney, Roberto and Kinder, Johannes and Cavallaro, Lorenzo},
  booktitle={28th USENIX security symposium (USENIX Security 19)},
  pages={729--746},
  year={2019}
}

@inproceedings{chen2023continuous,
  title={Continuous learning for android malware detection},
  author={Chen, Yizheng and Ding, Zhoujie and Wagner, David},
  booktitle={32nd USENIX Security Symposium (USENIX Security 23)},
  pages={1127--1144},
  year={2023}
}

@inproceedings{ling2024wolf,
  title={A Wolf in Sheep's Clothing: Practical Black-box Adversarial Attacks for Evading Learning-based Windows Malware Detection in the Wild},
  author={Ling, Xiang and Wu, Zhiyu and Wang, Bin and Deng, Wei and Wu, Jingzheng and Ji, Shouling and Luo, Tianyue and Wu, Yanjun},
  booktitle={33rd USENIX Security Symposium (USENIX Security 24)},
  pages={7393--7410},
  year={2024},
  organization={USENIX Association}
}

@inproceedings{lucas2023adversarial,
  title={Adversarial training for $\{$Raw-Binary$\}$ malware classifiers},
  author={Lucas, Keane and Pai, Samruddhi and Lin, Weiran and Bauer, Lujo and Reiter, Michael K and Sharif, Mahmood},
  booktitle={32nd USENIX Security Symposium (USENIX Security 23)},
  pages={1163--1180},
  year={2023}
}

@inproceedings{li2025revisiting,
  title={Revisiting Concept Drift in Windows Malware Detection: Adaptation to Real Drifted Malware with Minimal Samples},
  author={Li, Adrian Shuai and Iyengar, Arun and Kundu, Ashish and Bertino, Elisa},
  year={2025},
  organization={Network and Distributed System Security (NDSS) Symposium}
}

@misc{malwarebazaar_api,
  title        = {{MalwareBazaar API}},
  author       = {Abuse.ch},
  howpublished = {\url{https://bazaar.abuse.ch/api/}},
  year         = {2025},
  note         = {Online; accessed 29-May-2025}
}

@inproceedings{tripathi2025towards,
  title={Towards Explainable Drift Detection and Early Retrain in ML-Based Malware Detection Pipelines},
  author={Tripathi, Jayesh and Gomes, Heitor and Botacin, Marcus},
  booktitle={International Conference on Detection of Intrusions and Malware, and Vulnerability Assessment},
  pages={3--24},
  year={2025},
  organization={Springer}
}

@inproceedings{abusnaina2025exposing,
  title={Exposing the Limitations of Machine Learning for Malware Detection Under Concept Drift},
  author={Abusnaina, Ahmed and Anwar, Afsah and Saad, Muhammad and Alabduljabbar, Abdulrahman and Jang, Rhongho and Salem, Saeed and Mohaisen, David},
  booktitle={International Conference on Web Information Systems Engineering},
  pages={273--289},
  year={2025},
  organization={Springer}
}

@article{botacin2025towards,
  title={Towards more realistic evaluations: The impact of label delays in malware detection pipelines},
  author={Botacin, Marcus and Gomes, Heitor},
  journal={Computers \& Security},
  volume={148},
  pages={104122},
  year={2025},
  publisher={Elsevier}
}

@article{li2025lfreeda,
  title={LFreeDA: Label-Free Drift Adaptation for Windows Malware Detection},
  author={Li, Adrian Shuai and Bertino, Elisa},
  journal={arXiv preprint arXiv:2511.14963},
  year={2025}
}

@inproceedings{wang2025dart,
  title={Dart: A principled approach to adversarially robust unsupervised domain adaptation},
  author={Wang, Yunjuan and Hazimeh, Hussein and Ponomareva, Natalia and Kurakin, Alexey and Hammoud, Ibrahim and Arora, Raman},
  booktitle={2025 IEEE Conference on Secure and Trustworthy Machine Learning (SaTML)},
  pages={773--796},
  year={2025},
  organization={IEEE}
}

@article{zhang2022semantics,
  title={Semantics-preserving reinforcement learning attack against graph neural networks for malware detection},
  author={Zhang, Lan and Liu, Peng and Choi, Yoon-Ho and Chen, Ping},
  journal={IEEE Transactions on Dependable and Secure Computing},
  volume={20},
  number={2},
  pages={1390--1402},
  year={2022},
  publisher={IEEE}
}

@inproceedings{lucas2021malware,
  title={Malware makeover: Breaking ML-based static analysis by modifying executable bytes},
  author={Lucas, Keane and Sharif, Mahmood and Bauer, Lujo and Reiter, Michael K and Shintre, Saurabh},
  booktitle={Proceedings of the 2021 ACM Asia Conference on Computer and Communications Security},
  pages={744--758},
  year={2021}
}

@inproceedings{lucas2024training,
  title={Training robust ml-based raw-binary malware detectors in hours, not months},
  author={Lucas, Keane and Lin, Weiran and Bauer, Lujo and Reiter, Michael K and Sharif, Mahmood},
  booktitle={Proceedings of the 2024 on ACM SIGSAC Conference on Computer and Communications Security},
  pages={124--138},
  year={2024}
}

@inproceedings{kreuk2018adversarial,
  title={Adversarial examples on discrete sequences for beating whole-binary malware detection},
  author={Kreuk, Felix and Barak, Assi and Aviv-Reuven, Shir and Baruch, Moran and Pinkas, Benny and Keshet, Joseph},
  booktitle={Proc. NeurIPSW},
  year={2018}
}

@article{madry2017towards,
  title={Towards deep learning models resistant to adversarial attacks},
  author={Madry, Aleksander and Makelov, Aleksandar and Schmidt, Ludwig and Tsipras, Dimitris and Vladu, Adrian},
  journal={arXiv preprint arXiv:1706.06083},
  year={2017}
}

@inproceedings{lo2022exploring,
  title={Exploring adversarially robust training for unsupervised domain adaptation},
  author={Lo, Shao-Yuan and Patel, Vishal},
  booktitle={Proceedings of the Asian Conference on Computer Vision},
  pages={4093--4109},
  year={2022}
}

@inproceedings{zhu2023srouda,
  title={Srouda: meta self-training for robust unsupervised domain adaptation},
  author={Zhu, Wanqing and Yin, Jia-Li and Chen, Bo-Hao and Liu, Ximeng},
  booktitle={Proceedings of the AAAI Conference on Artificial Intelligence},
  volume={37},
  number={3},
  pages={3852--3860},
  year={2023}
}

@article{szegedy2013intriguing,
  title={Intriguing properties of neural networks},
  author={Szegedy, Christian and Zaremba, Wojciech and Sutskever, Ilya and Bruna, Joan and Erhan, Dumitru and Goodfellow, Ian and Fergus, Rob},
  journal={arXiv preprint arXiv:1312.6199},
  year={2013}
}

@article{goodfellow2014explaining,
  title={Explaining and harnessing adversarial examples},
  author={Goodfellow, Ian J and Shlens, Jonathon and Szegedy, Christian},
  journal={arXiv preprint arXiv:1412.6572},
  year={2014}
}

@article{chakraborty2018adversarial,
  title={Adversarial attacks and defences: A survey},
  author={Chakraborty, Anirban and Alam, Manaar and Dey, Vishal and Chattopadhyay, Anupam and Mukhopadhyay, Debdeep},
  journal={arXiv preprint arXiv:1810.00069},
  year={2018}
}

@article{hendrycks2019benchmarking,
  title={Benchmarking neural network robustness to common corruptions and perturbations},
  author={Hendrycks, Dan and Dietterich, Thomas},
  journal={arXiv preprint arXiv:1903.12261},
  year={2019}
}

@article{ren2020adversarial,
  title={Adversarial attacks and defenses in deep learning},
  author={Ren, Kui and Zheng, Tianhang and Qin, Zhan and Liu, Xue},
  journal={Engineering},
  year={2020}
}

@misc{domainrobust,
  title={DomainRobust: A testbed for adversarial robustness of domain adaptation},
  author={Wang, Haohan and Ge, Songwei and Xing, Eric P. and Lipton, Zachary C.},
  year={2025},
  howpublished={\url{https://github.com/google-research/domain-robust}},
  note={Our DART implementation is based on this codebase}
}

@inproceedings{digregorio2024tarallo,
  title={Tarallo: Evading behavioral malware detectors in the problem space},
  author={Digregorio, Gabriele and Maccarrone, Salvatore and D’Onghia, Mario and Gallo, Luigi and Carminati, Michele and Polino, Mario and Zanero, Stefano},
  booktitle={International Conference on Detection of Intrusions and Malware, and Vulnerability Assessment},
  pages={128--149},
  year={2024},
  organization={Springer}
}

@inproceedings{d2023lookin,
  title={Lookin'Out My Backdoor! Investigating Backdooring Attacks Against DL-driven Malware Detectors},
  author={D'Onghia, Mario and Di Cesare, Federico and Gallo, Luigi and Carminati, Michele and Polino, Mario and Zanero, Stefano},
  booktitle={Proceedings of the 16th ACM Workshop on Artificial Intelligence and Security},
  pages={209--220},
  year={2023}
}

@article{ceschin2023fast,
  title={Fast \& furious: On the modelling of malware detection as an evolving data stream},
  author={Ceschin, Fabr{\'\i}cio and Botacin, Marcus and Gomes, Heitor Murilo and Pinag{\'e}, Felipe and Oliveira, Luiz S and Gr{\'e}gio, Andr{\'e}},
  journal={Expert Systems with Applications},
  volume={212},
  pages={118590},
  year={2023},
  publisher={Elsevier}
}

@article{ceschin2024machine,
  title={Machine learning (in) security: A stream of problems},
  author={Ceschin, Fabr{\'\i}cio and Botacin, Marcus and Bifet, Albert and Pfahringer, Bernhard and Oliveira, Luiz S and Gomes, Heitor Murilo and Gr{\'e}gio, Andr{\'e}},
  journal={Digital Threats: Research and Practice},
  volume={5},
  number={1},
  pages={1--32},
  year={2024},
  publisher={ACM New York, NY}
}

@article{d2020dissection,
  title={On the dissection of evasive malware},
  author={D’Elia, Daniele Cono and Coppa, Emilio and Palmaro, Federico and Cavallaro, Lorenzo},
  journal={IEEE Transactions on Information Forensics and Security},
  volume={15},
  pages={2750--2765},
  year={2020},
  publisher={IEEE}
}

@inproceedings{he2025combating,
  title={Combating concept drift with explanatory detection and adaptation for android malware classification},
  author={He, Yiling and Lei, Junchi and Qin, Zhan and Ren, Kui and Chen, Chun},
  booktitle={Proceedings of the 2025 ACM SIGSAC Conference on Computer and Communications Security},
  pages={978--992},
  year={2025}
}

@article{galloro2022systematical,
  title={A systematical and longitudinal study of evasive behaviors in windows malware},
  author={Galloro, Nicola and Polino, Mario and Carminati, Michele and Continella, Andrea and Zanero, Stefano},
  journal={Computers \& security},
  volume={113},
  pages={102550},
  year={2022},
  publisher={Elsevier}
}

\appendix
\section{Additional Tables}\label{sec:appendix}

\begin{table}[htbp]
\caption{Malware-to-benign ratios per testing month.}
\label{mb24_ratio}
\centering
\small
\begin{tabular}{@{}crrr@{}}
\toprule
\textbf{\begin{tabular}[c]{@{}c@{}}Test\\Month\end{tabular}} & \textbf{\begin{tabular}[c]{@{}c@{}}Source\\Tr./Te.\end{tabular}} & \textbf{\begin{tabular}[c]{@{}c@{}}Target\\Train\end{tabular}} & \textbf{\begin{tabular}[c]{@{}c@{}}Target\\Test\end{tabular}} \\ \midrule
08/2024 & 0.6:1 & 0.5:1 & 0.6:1 \\
09/2024 & 0.6:1 & 0.6:1 & 0.5:1 \\
10/2024 & 0.6:1 & 0.5:1 & 0.5:1 \\
11/2024 & 0.6:1 & 0.5:1 & 0.4:1 \\
12/2024 & 0.6:1 & 0.4:1 & 0.4:1 \\
\textbf{Avg} & \textbf{0.6:1} & \textbf{0.5:1} & \textbf{0.5:1} \\ \bottomrule
\end{tabular}
\end{table}

\begin{table}[htbp]
\caption{Candidate configurations for tuning the AdvDA model. We evaluated $10$ configurations and selected the one with the highest accuracy on the source test set, shown in bold.}
\label{tab:hyperparam}
\centering
\begin{tabular}{lccccccc}
\toprule
\textbf{Config} & \textbf{Backbone} & \textbf{LR} & \textbf{$\lambda_2$} & \textbf{$\lambda_1$} & \textbf{Batch} & \textbf{Epochs} & \textbf{Options} \\
\midrule
1         & CNN      & $10^{-3}$ & 0.1  & 0.1 & 32 & 30 & - \\
2  & CNN      & $10^{-3}$ & 0.1  & 0.5 & 32 & 30 & - \\
3 & CNN      & $10^{-3}$ & 0.2  & 0.1 & 32 & 30 & - \\
4        & CNN      & $10^{-3}$ & 0.15 & 0.3 & 32 & 30 & - \\
5     & CNN      & $10^{-4}$ & 0.1  & 0.2 & 32 & 50 & - \\
6     & CNN      & $10^{-3}$ & 0.1  & 0.2 & 64 & 30 & - \\
7   & CNN      & $10^{-3}$ & 0.1  & 0.5 & 32 & 30 & class weight. \\
8        & CNN      & $10^{-3}$ & 0.1  & 0.5 & 32 & 30 & grad clip \\
9        & Deep CNN & $10^{-3}$ & 0.1  & 0.5 & 32 & 30 & - \\
\textbf{10} & \textbf{ResNet-18} & $\mathbf{10^{-4}}$ & \textbf{0.02} & \textbf{0.5} & \textbf{32} & \textbf{30} & \textbf{grad clip} \\
\bottomrule
\end{tabular}
\end{table}

\end{document}